\documentclass[aps,prl,twocolumn]{revtex4}

\usepackage{graphicx}% Include figure files
\usepackage{dcolumn}% Align table columns on decimal point
\usepackage{bm}% bold math
\usepackage[normalem]{ulem}

\begin{document}

\title{$^4$He on a single graphene sheet
}  

\author{M.C. Gordillo}
\affiliation{Departamento de Sistemas F\'{\i}sicos, Qu\'{\i}micos 
y Naturales. Facultad de Ciencias Experimentales. Universidad Pablo de Olavide. 
Carretera de Utrera, km 1. 41013 Sevilla. Spain.}

\author{J. Boronat}
\affiliation{Departament de F\'{\i}sica i Enginyeria Nuclear, 
Universitat Polit\`ecnica de Catalunya, 
B4-B5 Campus Nord. 08034 Barcelona. Catalonia. Spain.}

\date{\today}

\begin{abstract}
The phase diagram of the first layer of $^4$He adsorbed on a single
graphene sheet has been calculated by a series of 
diffusion Monte Carlo calculations including corrugation effects. 
As the number of C-He interactions is reduced with respect to graphite 
the binding
energy of  $^4$He atoms to graphene is approximately 13.4
K/He atom smaller. 
Our results indicate that the phase diagram is 
qualitatively similar
to that of  helium on top of graphite.
A two-dimensional liquid film on graphene is predicted to be metastable with respect to
the commensurate solid but the difference in energy between both phases is
very small, opening the
possibility of such a liquid film to be experimentally observed.
\end{abstract}

%\keywords{}

\maketitle

Graphite is a a well known form of carbon, made of   
two dimensional carbon layers 
glued to each other by interactions of dispersion type in the $z$ 
direction, and separated by a distance of 3.35 \AA. 
Within each of those two dimensional layers the
carbon atoms are located in the nodes of a honeycomb lattice, 
each of them being bound to three others by 
covalent interactions.     
Even though it is well known that to exfoliate graphite is relatively easy,
it was only recently reported the isolation of a single and stable 
two-dimensional sheet of carbon by mechanical cleavage  
\cite{science2004,pnas2005}. This structure is termed graphene and 
it has been predicted
to be unstable since the thermal fluctuations
would make the crystal structure collapse \cite{Landau}. However, 
the experiments
show that at least it is kinetically 
stable \cite{pnas2005}. This novel singular 
material has already attracted the attention of the   
scientific community, basically for its novel electrical properties
\cite{nature1,nature2,jpcbgra,natmat}. 

In this work, we are interested in graphene as a new adsorber. 
Since graphite is a set of graphene layers, one would expect
a difference in the binding energy of the adsorbed species
that could lead to a change in their phase diagram. 
We have carried out this analysis for $^4$He in the limit of
zero temperature, and studying the differences between graphene and 
graphite. 
To this end, we have performed Diffusion Monte 
Carlo simulations of a system of $^4$He atoms
adsorbed on top of 
a variable number of graphene layers, ranging from one to eight. This last
number was found to be an acceptable model for graphite, since 
 all the properties calculated were similar for eight and nine layers
within the error bars obtained from the simulation data. The layers were
supposed to be parallel to each other, separated  
by the typical graphite distance, and stacked in the A-B-A-B way characteristic
of this compound. The helium densities were kept within the limits of 
a first layer ( $<$  0.12 \AA$^{-2}$ \cite{grey}). 

We used Diffusion Monte Carlo (DMC) 
because
it is able to obtain the true ground state for bosonic systems, such as 
a set of $^4$He particles \cite{boro94} in the limit of 
0 K. 
However, for the technique to work we have to provide 
a reasonable approximation for the ground-state  wave function, what it is 
called  the trial function, that
collects all the information known {\em a priori} about the system.
In this work, we used as a first trial wave function  
\begin{equation}
\Phi({\bf r_1, r_2, ..., r_N}) = \prod_{i<j} \exp \left[-\frac{1}{2} 
\left(\frac{b_{\rm He-He}}{r_{ij}} \right)^5 \right] \prod_i \Psi(z_i) \ ,
\label{trial1}
\end{equation} 
that depends on the coordinates of the helium atoms 
$r_1, r_2, ..., r_N$, and where the first term is the usual 
Jastrow function depending on the
helium interatomic distances $r_{ij}$, with $b_{\rm He-He} = 3.07$ \AA 
\cite{boro94}. For the second part of the
wave function ($\Psi(z)$), we followed
Withlock and collaborators \cite{Whitlock}, and solved the 
one-dimensional Schroedinger equation describing a single helium atom 
moving along the axis perpendicular to the graphene layers ($z$) for
an averaged C-He potential that neglected corrugation. 
The one-body ground state wave function obtained, 
$\Psi(z)$, is displayed in Fig. 1.  
The
He-He potential was taken from Ref. \onlinecite{aziz} while the individual
C-He interactions were assumed to be of 
Lennard-Jones type 
\cite{cole}. This means that in
our many-body 
calculations the effects of carbon corrugation in the C-He interaction 
are fully considered. 
 
\begin{figure}
\setlength\fboxsep{0pt} \centering
\scalebox{1.00}{\includegraphics[width=0.8\linewidth]{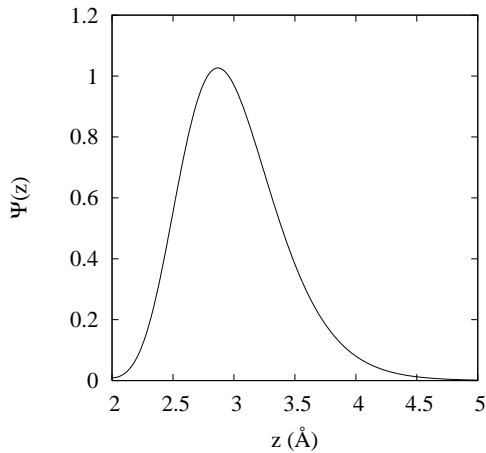}}
\caption{Solution for the one-body Schroedinger equation describing a single
$^4$He atom under an averaged C-He interaction in the $x,y$ plane. 
$z$ indicates distance to such plane, located at $z$ = 0.  
}
\label{fig1}
\end{figure}

Obviously, the former trial function (\ref{trial1}) is an adequate representation 
for a system with translational invariance, i.e., a liquid. To do 
simulations
for a solid, we multiply the previous $\Phi({\bf r_1, r_2,
..., r_N})$ (\ref{trial1}) by a term 
 of the form 
\begin{equation} 
\prod_i \exp{-a [(x_i-x_{\rm site})^2 + (y_i-y_{\rm site})^2]} \ ,
\label{trial2}
\end{equation}
where $x_{\rm site},y_{\rm site}$ are the coordinates of the 
crystallographical positions
around which the $^4$He atoms are localized and $a$ is a 
constant variationally
optimized. These positions were different for
each of the solid phases considered: a commensurate 
$\sqrt3 \times \sqrt3$, of surface density 
$0.0636$ \AA$^{-2}$ \cite{colebook}, a commensurate structure 
(7/16) reported by 
Corboz and coworkers \cite{boninx} ($\rho = 0.0835$ \AA$^{-2}$), and
three commensurate structures suggested to appear in the 
experimental phase diagram of graphite in Ref. \onlinecite{grey2} 
(2/5, $\rho = 0.0763$ \AA$^{-2}$; 3/7, $\rho = 0.0818$ \AA$^{-2}$ ;
31/75 $\rho = 0.0763$ \AA$^{-2}$)  . We also considered 
several triangular incommensurate solids of different densities, obtained by 
varying the helium-helium 
distance in the $x,y$ plane on the (outer) graphene sheet. The
optimal values of the parameter $a$ in Eq. (\ref{trial2}) are $a = 0.31$ 
\AA$^{-2}$ for all the commensurate structures, and ranges 
from $a = 0.15$ to $0.77$ \AA$^{-2}$
in the case
of the incommensurate triangular structures.
The quality of the different trial functions to describe the
system can be ascertained by looking at the energy variances of every
single calculation, in all cases of the order of 0.01-0.02 K.   

\begin{figure}
\setlength\fboxsep{0pt} \centering
\scalebox{1.00}{\includegraphics[width=0.8\linewidth]{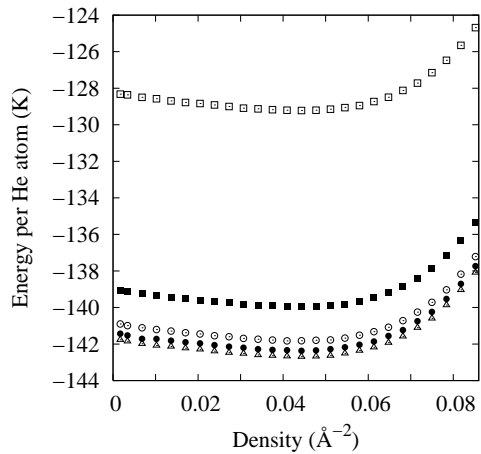}}
\caption{Energy per atom versus the density for a liquid phase of helium
atoms on top of one or several graphene sheets. Open squares, $n$ = 1;
full squares, $n$ = 2; open circles, $n$ = 3; full circles, $n$=4; 
open triangles, $n$ = 8. 
}
\label{fig2}
\end{figure}

Fig. 2 displays the energy per $^4$He atom as a function of the helium density
for the liquid phase and for
 the number of graphene layers ($n$) 
considered. 
The corresponding densities were obtained by varying the number of helium 
atoms on top of a fixed simulation cell, a rectangle of $34.43 \times
34.08$ \AA$^2$. The use of a simulation cell with different size or form 
did not change the results.
We can see that there is a very small but significant difference between the 
binding energy values for $n$ = 4 and $n$ = 8 ($-142.37 \pm 0.01$ K versus
$-142.69 \pm 0.01$ K), for the density corresponding to the minimum energy,
$0.044$ \AA$^{-2}$. This equilibrium density is nearly
equal to the one for purely two-dimensional liquid $^4$He ($0.043$
\AA$^{-2}$)~\cite{giorgini}.  
The energy differences between structures with the
same densities for eight and nine carbon 
sheets are below the corresponding
error bars
and therefore $n=8$ can be considered the graphite limit.
 We also found that the curves are  
 similar to each other, up to the point to 
nearly collapse
in one single function when the differences in the infinite dilution limit
are corrected. The
energies of the liquid phase at the equilibrium density are 
shown in Table I. The infinite-dilution energies 
were obtained from fittings to a third-degree polynomials in the density
range $\rho < 0.02$ \AA$^{-2}$. The binding energy
of a single atom for 
$n  = 8$ is fully compatible
with the experimental data \cite{exp1,elgin}, and slightly different
from 
the Path Integral Monte Carlo calculations of Ref. \onlinecite{manousakis2} 
(-143.09 $\pm$ 0.27 K versus -141.64 $\pm$ 0.01 K for
the present work), probably
because of the different C-He interactions used.

\begin{table*}
\caption{Energy per atom, in K, for several helium arrangements. $n$
indicates
the number of graphene layers considered. See further explanation in the 
text. 
}
\begin{tabular}{ccccc} \hline
 n  & Infinite dilution & Liquid & $\sqrt 3 
 \times \sqrt 3$ & Incommensurate \\ \hline 
1  & -128.26 $\pm$ 0.04  &   -129.221 $\pm$ 0.009& -129.282 $\pm$ 0.007 & 
 -126.6 $\pm$ 0.2  \\
2 & -139.02 $\pm$ 0.01 & -139.96 $\pm$ 0.01 &  -140.067  $\pm$ 0.009 & 
-137.3 $\pm$ 0.2  \\
4 & -141.24 $\pm$ 0.09 & -142.37 $\pm$ 0.01 &  -142.45  $\pm$ 0.01 & 
 -139.7 $\pm$ 0.2  \\
8 & -141.64 $\pm$ 0.03 & -142.69 $\pm$ 0.01 &  -142.81  $\pm$ 0.01 & 
-140.0 $\pm$ 0.2  \\
 \hline
\end{tabular}
\end{table*}

In the scenario described above 
the liquid is not the most stable phase 
at $T = 0$ K. The 
energies of the liquid phase at
equilibrium for different values of $n$ are given in
Table I, and can be compared with the binding
energy for the $\sqrt3 \times \sqrt3$
commensurate structure.
The simulation of the $\sqrt3 \times \sqrt3$ solid phase has been
performed using 120 atoms in a 
$44.27 \times 42.60$ \AA$^2$ cell. 
In the Table, we can see that in all cases, commensurate
solids are more bound, but admittedly for 
very small margins.  The ground state of
 $^4$He on top of
any graphene compound is the $\sqrt3 \times \sqrt3$ registered phase.
This result is stable within plausible uncertainties ($\sim 5$\%) of the C-He
interaction energy scale ($\varepsilon$): we have verified that 
the liquid would appear as the stable phase by decreasing much more the
energy $\varepsilon$ (20 \%).
Our description of the phase diagram agrees with the Path Integral 
Monte Carlo one of 
Pierce and Manousakis \cite{manousakis1, manousakis2} for graphite.  
However, the small difference between the energies of the liquid and 
commensurate phases makes also plausible the scenario given by Greywal
and Busch \cite{grey}, with a liquid phase of density 
$\sim 0.04$ \AA$^{-2}$ as
a very close metastable state. In the densities between two stable structures in a 
phase diagram, the system divides itself in patches of the coexisting phases.
Those patches could be puddles of liquid or clusters of the 
 $\sqrt3 \times \sqrt3$ structure \cite{manousakis2,
manousakis1} surrounded by empty space. 
 
\begin{figure}
\setlength\fboxsep{0pt} \centering
\scalebox{1.00}{\includegraphics[width=0.8\linewidth]{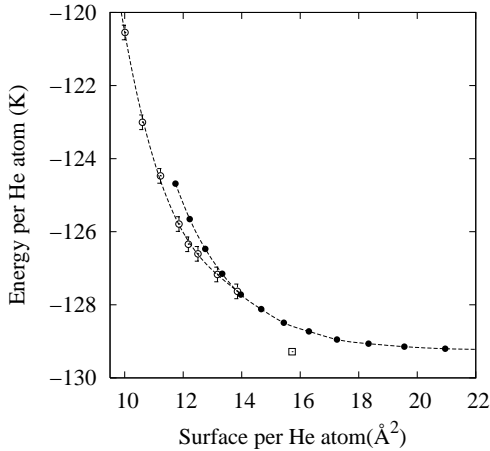}}
\caption{Energy versus the inverse of the density for helium on top of
a single graphene layer. Full circles, liquid phase; open circles, 
incommensurate triangular lattice; open square, registered $\sqrt 3 \times 
\sqrt 3 $ phase.  Where not
displayed, error bars are of the same size or smaller than the symbols.
The dashed line is the result of a third order polynomial fit to the
incommensurate results. 
}
\label{fig3}
\end{figure}

The experimental phase diagram of the first layer of $^4$He on top of graphite
indicates that at high enough density, the stable phase is an incommensurate
triangular phase \cite{grey,grey2}. In between the   
$\sqrt3 \times \sqrt3$ registered phase and the
incommensurate one,  
different phases have been suggested \cite{grey2,boninx, manousakis2}, 
from a domain wall phase (DWP) to 
several commensurate structures of different densities. We performed
DMC calculations to check the stability of these commensurate
phases on top of graphene versus a triangular arrangement of the same 
density. The results are displayed in Table II. The DWP was not checked
because in a DMC calculation the atoms in a wall have to be located in definite
positions. Those positions have to be arbitrary (for instance, crossing the 
simulation cell diagonally or vertically, or in arrays of three atoms 
instead of two) since no experimental information
is available. This means that the simulation results do not
represent the DWP phase meaningfully, only giving the corresponding 
energies of a particular atom arrangement including a wall. 
Table II indicates that all the commensurate structures have energies
per atom larger than the corresponding triangular phase of the same
density. However, and except in the case of the 2/5 phase, the differences
are within two error bars. This means that they could be metastable states
that could be  experimentally observed.    

\begin{table*}
\caption{Energy per atom, in K, for several commensurate helium structures.
The results labeled "incommensurate" indicate the energies for a triangular
phase of the same density as the one in the previous line. }
\begin{tabular}{ccccc} \hline
 Compound   & $\frac{2}{5}$  &  $\frac{31}{75}$ & $\frac{3}{7}$ 
 &  $\frac{7}{16}$ \\ \hline 
Graphene  & -125.81 $\pm$ 0.01  &  -126.50 $\pm$ 0.02 & -126.07 $\pm$ 0.01 & 
 -125.89 $\pm$ 0.01  \\
Graphene (incommensurate) & -127 $\pm$ 0.2  &   -126.8 $\pm$ 0.2& -126.3 $\pm$ 0.2 & 
 -126.0 $\pm$ 0.2  \\
Graphite  & -139.25 $\pm$ 0.01 & -139.96 $\pm$ 0.01 &  -139.54  $\pm$ 0.01 & 
-139.33 $\pm$ 0.01  \\
Graphite (incommensurate) & -140.5 $\pm$ 0.2 & -140.2 $\pm$ 0.2 &  -139.7  $\pm$ 0.2 & 
-140.0 $\pm$ 0.2  \\
Graphite (+ McLachlan)  & -138.75 $\pm$ 0.03 & -139.50 $\pm$ 0.02 &  -139.02  $\pm$ 0.01 & 
-138.81 $\pm$ 0.01  \\
Graphite (+ McLachlan incommensurate) & -140.1 $\pm$ 0.2 & -139.6 $\pm$ 0.2 &  -139.3  $\pm$ 0.2 & 
-138.9 $\pm$ 0.2  \\
 \hline
\end{tabular}
\end{table*}

The phase diagram of $^4$He on top of a single graphene 
layer can
be established with the help of Fig. 3. There, we show 
the energy per $^4$He
atom as a function of the inverse of the surface density. 
The error bars of the incommensurate structure are
noticeable larger than in the other cases, because  
they are not simply the statistical errors of the simulated energies, 
but the result of averaging over different positions of the helium 
crystallographical sites on top of the graphene layer. 
For each point in the figure, we performed four
different calculations considering similar triangular helium lattices, 
but displaced a little with respect to each other and averaged the
energy results. Thus, we take 
into account the incommensurability of solid
$^4$He on top of graphene.  
The only stable commensurate solid, should be in
equilibrium with a triangular incommensurate structure. The limits
of the coexistence phase should have to be determined by a 
double-tangent Maxwell construction. Since the commensurate phase is defined
by a single density, we can only approximate the result, by drawing 
a line from this point to the one in the triangular lattice
energy
curve that would result in the smallest pressure. This imperfect solution
gives us that the $\sqrt3 \times \sqrt3$ structure is in equilibrium with
a triangular phase of $\sim 0.08$ \AA$^{-2}$, in agreement with
experimental data for graphite \cite{grey}. In between both equilibrium
densities, there would be a coexistence zone formed by patches of 
both solids \cite{manousakis2}, forming a DWP. This entire picture for
the phase diagram in the $n = 1$ case is common to $n=2$,4,8. The 
energies per helium atom of the $\sqrt3 \times \sqrt3$ phase and the 
triangular one at $0.08$ \AA$^{-2}$ (equilibrium density), are given
in Table I (Incommensurate). The main  
 difference between graphene and
graphite is then an offset of $\sim$ 13.4 K in the binding energies of
$^4$He of the different compounds.

All the above results for graphene and graphite were calculated 
without taking into account any three body C-He interaction.  
To do so, one can introduce the so-called McLachlan interaction, 
between a carbon substrate represented by a semi-infinite slab and 
a couple of helium atoms on top of it \cite{bruch}. This means that
the McLachlan term only can be applied meaningfully to graphite.
The model of graphite used in our calculations
is not a slab but a stack of graphene sheets, 
implying that
the use of this term is only an approximation.  
In previous calculations \cite{got, manousakis1,manousakis2},
this term was found to favor the liquid versus the commensurate
$\sqrt 3 \times \sqrt 3$ structure. To check its influence on the
energy of the system, we performed DMC calculations on the $\sqrt 3 \times \sqrt 3$
and liquid phases on top of graphite in the very same conditions 
given above. The binding energy for the commensurate solid was
-142.44 $\pm$ 0.01 K versus -142.81 $\pm$ 0.01 K of Table I, i.e.
a difference of 0.37 K. The density minimum for the liquid 
phase changed from 0.044 \AA$^{-2}$ to 0.041 \AA$^{-2}$, with a
binding energy of -142.49 $\pm$ 0.01 K versus the -142.69 $\pm$ 0.01 K
of Table I. This would mean that the liquid is the stable phase
for $^4$He on top of graphite, but with a difference even smaller
than in the above calculations. We also repeated the calculations
including this term for all the registered phases suggested above. 
The results are given in Table II. We found that the McLachlan interaction
change the binding energies but not the fact that they are metastable with
respect to the incommensurate triangular phase. 
The results are not applicable to
graphene, since the model of a thick carbon slab does not apply to 
that substrate.

Summarizing, we have performed diffusion Monte Carlo calculations 
of $^4$He adsorbed on graphene for the first time using an accurate He-He 
interatomic
potential. Our results show that the ground state corresponds to a 
 $\sqrt3 \times \sqrt3$ commensurate solid. However, the 
difference
 in energy between a metastable liquid film and the solid is very
 tiny opening  the
 possibility of observing experimentally a two-dimensional superfluid
 liquid phase. 
 Other commensurate phases have also been studied and
 found to be metastable, both in graphene and graphite.
 The phase diagram of $^4$He on graphene is qualitatively equal to the one
 of $^4$He on graphite that we have obtained by adding graphene sheets up
  to $n =8$.

%\begin{acknowledgments}
We acknowledge partial financial support from 
The Spanish Ministry of Education and Science (MEC), grants  
 FIS2006-02356 and  FIS2005-04181, Junta de Andaluc\'{\i}a, grant 
P06-FQM-01869, and Generalitat de Catalunya, grant  2005GR-00779.
%\end{acknowledgments}

\end{document}